\documentclass[12pt]{article}
\pdfoutput=1
\usepackage{graphicx}
\usepackage{subcaption}
\usepackage{amsfonts}
\usepackage{amssymb,amsmath}
\usepackage{hyperref}
\usepackage{comment}
\usepackage{multicol}
\usepackage[usenames,dvipsnames]{xcolor}
\definecolor{rosso}{cmyk}{0,1,1,0.3}
\definecolor{verde}{cmyk}{0.8,0,0.6,0.25}
\definecolor{bluc}{cmyk}{1,0.4,0,0.1}
\definecolor{blucc}{cmyk}{0.8,0.3,0,0}

\def\mdm{m_{\rm DM}}
\def\tdm{\tau_{\rm DM}}
\def\GEV{\rm GeV}
\def\KEV{\rm keV}
\def\ax{\tilde{a}}

\newcommand{\lrfp}[3]{ \left(\frac{#1}{#2}\right)^{#3} }

\begin{document}
\begin{titlepage}
\begin{flushright}
UT-14-10\\
\end{flushright}

\vskip 3cm

\begin{center}

{\Large\bf Axino dark matter in light of an anomalous X-ray line}

\vskip .5in

{
Seng Pei Liew }

\vskip .3in

{\em
Department of Physics, University of Tokyo, Bunkyo-ku, Tokyo 113-0033, Japan}

\begin{abstract}
  Axino as the superpartner of axion that solves the strong CP problem can be a good candidate of dark matter. Inspired by the 3.5 keV X-ray line signal found to be originated from galaxy clusters and Andromeda galaxy, we study axino models with R-parity violations, and point out that axino dark matter with trilinear R-parity violations is an attractive scenario that reproduces the X-ray line. The Peccei-Quinn scale is required to be $f_a\sim{\cal O}(10^{9}-10^{11})~\GEV$ for trilinear R-parity violating couplings $\lambda\sim {\cal O} (10^{-3}-10^{-1})$ in order to explain the line signal. Moreover, the right-handed stau is predicted to be light, i.e.~$\sim{\cal O}(100)$ GeV, and thus can be looked for at the LHC. Cosmological aspects of the model are also discussed in this study. 
\end{abstract}

\end{center}
\end{titlepage}


\section{Introduction}
 Axion is a product of the Peccei-Quinn (PQ) mechanism introduced to solve the strong CP problem~\cite{Kawasaki:2013ae}. In supersymmetric (SUSY) models of axion, $a$, axino, $\ax$ (saxion, $\sigma$) appears as the fermionic (scalar) partner of axion. Axino is neutral and if it is the lightest supersymmetric particle (LSP), it can be a candidate of dark matter (DM) which makes up 27 \% of the energy density of the universe~\cite{Choi:2013lwa}.
 
With the introduction of R-parity violating (RPV) terms into the SUSY models, the LSP is destabilized and starts decaying into Standard Model (SM) particles. If the lifetime of the LSP is similar or longer than the age of the universe, it still can play the role of DM. Furthermore, this scenario opens up an opportunity to observe the signature of DM decay.

This kind of signature might have been discovered by two independent groups looking for X-ray line emissions originated from galaxy clusters as well as Andromeda galaxy~\cite{Bulbul:2014sua,Boyarsky:2014jta}. It has been found that there is an excess of X-ray emissions at around 3.5 keV, which has no explanation with known physics. Slowly decaying DM of mass $\mdm \sim 7$ keV is a viable interpretation of the line signal, although one requires its confirmation from other observational experiments and astrophysical objects. If the line signal is to be interpreted as DM decaying into a photon, its lifetime is estimated to be $\tdm\sim 10^{28}$ sec.       
 
The X-ray line observations have generated interests in interpreting the line signal with axino DM~\cite{Kong:2014gea,Choi:2014tva}.~\footnote{Axino DM with RPV has also been studied, albeit in different contexts, in~\cite{Kim:2001sh,Hooper:2004qf,Chun:2006ss,Endo:2013si}. See also~\cite{Hasenkamp:2011xh}, where gravitino is the LSP and axino is the heavier SUSY particle.}~These studies have been focusing on bilinear RPV, where axino decays into a photon and a neutrino via gaugino mixing with neutrinos. It is shown in~\cite{Kong:2014gea} that the PQ scale, $f_a$ needs to be $\sim 10^8-10^9~\GEV$ in order to reproduce the X-ray line. However, when the supernova (SN 1987A) bound on axion is taken into account, where the following relation has to be satisfied~\cite{Raffelt:2006cw}:
\begin{equation}
f_a \gtrsim 4 \times 10^8~\GEV,
\end{equation}
the remaining viable parameter region becomes much constrained. In~\cite{Choi:2014tva}, $f_a$ is pushed to values $10^9-10^{11}~\GEV$ by requiring the bino mass to be $\lesssim 10~\GEV$. On the other hand, other neutralinos must be kept much heavier than the bino, which is technically possible but raises questions of how and why such hierarchy exists.

In the present work, we instead explore axino DM with trilinear RPV in light of the anomalous 3.5 keV X-ray line.   Axino with trilinear RPV can decay into a photon and a neutrino ($\ax \to \gamma + \nu$) via Feynman diagrams with a loop involving fermion and sfermion. As will be shown in the following sections, $f_a$ can be made larger than $10^9 \GEV$, evading various astrophysical bounds on axion while being consistent with the X-ray line from axino DM decay. Within this framework, sfermions are relatively light, and RPV couplings are predicted to be large, making this model testable using colliders. 

The rest of the paper is organized as follows. In Section~\ref{sc:fr}, we lay down the framework of our study, discussing SUSY axion models and RPV. In Section~\ref{sc:ax}, we study in detail decaying axino DM that explains the 3.5 keV X-ray line. Before we conclude, we describe cosmological and phenomenological consequences and implications of our model.
 
\section{Framework}
\label{sc:fr}
\subsection{Supersymmetric models of axion}
Let us begin with a brief description of SUSY models of axion. Broadly speaking, there are two kinds of ``invisible" axion models, KSVZ~\cite{Kim:1979if,Shifman:1979if} and DFSZ~\cite{Dine:1981rt,Zhitnitsky:1980tq}. In SUSY models, one introduces new PQ superfields $\Phi$'s and couples them to matter superfields carrying PQ charges. Then, the axion supermultiplet, $A$ arises when the $U(1)$ PQ symmetry is broken.

 In KSVZ models, SM particles do not carry PQ charges. There exist couplings of $\Phi$'s with PQ-charged extra heavy quark superfields $Q, \bar{Q}$ via the superpotential $W_{\rm PQ}= k \Phi Q \bar{Q}$, where $k$ is a coupling constant. In DFSZ models, PQ superfields couple to the Higgs superfields $H_u$ and $H_d$, and there is no direct coupling to the lepton and quark sectors.   

The common feature of these models is that the axino couples to gauginos and gauge bosons via the anomaly-induced terms
\begin{eqnarray}
	 \mathcal{L}_{\tilde{a}\lambda A}&=& i \frac{\alpha_Y C_Y}{16 \pi f_a}\bar{\tilde{a}}
	 \gamma_5[\gamma^{\mu},\gamma^{\nu}]\tilde{B}B_{\mu \nu} + i \frac{\alpha_W C_W}{16 \pi f_a}
	 \bar{\tilde{a}}\gamma_5[\gamma^{\mu},\gamma^{\nu}]\tilde{W}^aW^a_{\mu \nu}
	 \nonumber
	 \\
	 & &+ i \frac{\alpha_s }{16 \pi f_a}\bar{\tilde{a}}
	 \gamma_5[\gamma^{\mu},\gamma^{\nu}]\tilde{g}G_{\mu \nu},
	 \label{eq:anom} 
	 \end{eqnarray}
where $C_Y$ and $C_W$ are ${\cal O}(1)$ model-dependent coupling constants. This is the consequence of the broken PQ symmetry at the scale $f_a$.

We are especially interested in interaction terms that are related to the emission of photon from axino. The first two terms in Eq.~(\ref{eq:anom}) are responsible for such emission in models with bilinear RPV. As will be discussed in the following section, interactions between axino, fermion, and sfermion, i.e.
\begin{eqnarray}
\mathcal{L}_{\ax \tilde{f}f}= C_{\ax \tilde{f}_Lf_L}\tilde{f}_L^*\bar{\ax}P_Lf +  C_{\ax \tilde{f}_Rf_R}\tilde{f}_R^*\bar{\ax}P_Rf^c + {\rm h.c.}\ ,
\label{eq:gauge}
 \end{eqnarray}
 where $C_{\ax \tilde{f}f}$'s are dimensionless coupling constants, are important in models with trilinear RPV. For KSVZ models, there is no tree-level contribution to such interactions. These interactions are induced at the one-loop level through bino, $U(1)_Y$ gauge boson, and fermion~\cite{Covi:2001nw,Hooper:2004qf}:  
\begin{eqnarray}     
C_{\ax \tilde{f}f} \propto  \frac{\alpha_Y^2 C_Y^2}{\pi^2}\frac{M_1}{f_a}{\rm log}\left(\frac{f_a}{M_1}\right),
\label{eq:ksvz}
\end{eqnarray}
where $M_1$ is the bino mass parameter. For DFSZ models, $C_{\ax \tilde{f}f}$ arises at the tree level from the mixing of axino with Higgsino, which in turn, mixes with gauginos. The mixing of axino with Higgsino is $\sim v/f_a$, where $v$ is the Higgs vacuum expectation value (VEV). Hence, 
\begin{eqnarray}     
C_{\ax \tilde{f}f} \simeq  \hat{g} \frac{v}{f_a},
\label{eq:dfsz} 
\end{eqnarray}
where $\hat{g}$ is the gauge coupling constant suppressed by the Higgsino-gaugino mixing.

One can compare the strength of $C_{\ax \tilde{f}f}$'s for KSVZ and DFSZ models from Eqs.~(\ref{eq:ksvz})-(\ref{eq:dfsz}). $\hat{g}$ is typically of order $10^{-2}$, and for $M_1\sim v$, KSVZ's $C_{\ax \tilde{f}f}$'s are approximately smaller than DFSZ's by $10^{-2}$.

\subsection{R-parity violations}
The most general form of RPV is represented by the following superpotential~\cite{Barbier:2004ez}:
\begin{equation}
W= \lambda_{ijk} L_{i}L_{j}{E}_{k}+\lambda_{ijk} ^{\prime }L_{i}Q_{j}{D}_{k}
+\lambda_{ijk} ^{\prime \prime }{U}_{i}{D}_{j}{D}_{k}+\mu_i L_i H_u,  
\label{W_bi} 
\end{equation}
with the summation among the indices $i,j,k=1,2,3$ denoting the lepton and quark generations assumed implicitly. The first three terms correspond to trilinear RPV and the last one corresponds to bilinear RPV. $L_i$, $E_i$, $Q_i$, $D_i$, $U_i$ and $H_u$ are the usual matter superfields in the Minimal Supersymmetric Standard Model (MSSM). $\lambda_{ijk},\,\lambda_{ijk} ^{\prime },\,\lambda_{ijk} ^{\prime \prime }$\,are dimensionless coupling constants whereas the coupling constants $\mu_i$'s carry mass dimension one. SM gauge symmetries demand the indices $i$ and $j$ ($j$ and $k$) of $\lambda_{ijk}$ ($\lambda_{ijk} ^{\prime \prime }$) to be antisymmetric. Since the $UUD$ operator is irrelevant to our study, $\lambda_{ijk} ^{\prime \prime }=0$ is imposed by assuming baryon number conservation.  

Let us discuss bilinear RPV in more detail. $\mu_i L_i H_u$ can be rotated away by redefining $L_i$ and $H_d$ as $L_i' = L_i - \epsilon_i H_d$ and $H_d' = H_d +\epsilon_i L_i$ with $\epsilon_i \equiv \mu_i/\mu$,
where $\mu$ is the Higgsino mass parameter originated from the MSSM superpotential $\mu H_u H_d$. In general, SUSY-breaking soft terms cannot be rotated away along with the redefinition, and this leads to non-zero sneutrino VEVs, $\langle \tilde \nu_i\rangle$. The value of $\langle \tilde \nu_i\rangle$ signifies the strength of bilinear RPV, and it is often parametrized in terms of  $\kappa_i \equiv \langle \tilde \nu_i\rangle/v $.

\section{Axino dark matter with R-parity violations}
\label{sc:ax}
\subsection{Decay rate}
\begin{figure}
\centering
\begin{subfigure}[b]{0.4\textwidth}
 \includegraphics[width=7cm]{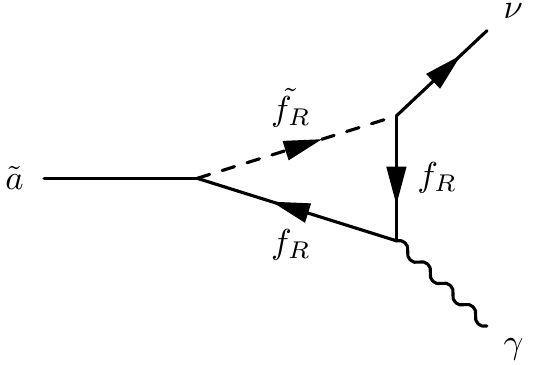}
 \caption{}
 \label{sf:a}
  \end{subfigure}
 \begin{subfigure}[b]{0.4\textwidth}
 \includegraphics[width=7cm]{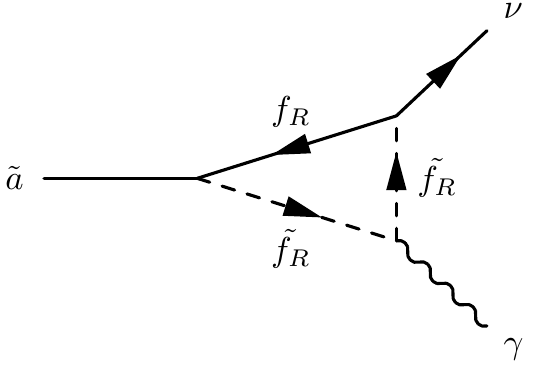}
 \caption{}
  \label{sf:b}
  \end{subfigure}
 \caption{Feynman diagrams of the process $\ax \to \gamma + \nu$ via $LLE$ or $LQD$ RPV operators. Here, right-handed (s)fermions are assumed to be running in the loop. The (s)fermion flow is in the opposite direction for the conjugate process $\ax \to \gamma + \bar{\nu}$.} 
\label{fig:feyn}
\end{figure}

We first calculate the one-loop radiative decay $\ax \to \gamma + \nu$ in theories with trilinear RPV induced by the operators $LLE$ and $LQD$.~\footnote{For a similar calculation that involves neutralino decaying radiatively via RPV operators, see~\cite{Mukhopadhyaya:1999gy}.} The relevant Feynman diagrams are shown in Figure~\ref{fig:feyn}.

 Several comments are in order before we present the final result. For the axino mass in consideration,  tree-level three-body decays are not allowed kinematically. Hence, the axino decays dominantly into $\gamma$ and $\nu$. Also, note that the (s)fermions running in the loop shown in Figure~\ref{fig:feyn} are charged (s)leptons for the $LLE$ scenario and down-type (s)quarks for the $LQD$ scenario. The one-loop decay amplitude involves a axino-sfermion-fermion vertex, with coupling constant given in Eqs.~(\ref{eq:ksvz})-(\ref{eq:dfsz}) for KSVZ and DFSZ models respectively. Notice that $C_{\ax \tilde{f}f}$ in KSVZ models is only an effective vertex induced at the one-loop level. A more rigorous calculation of $\ax \to \gamma + \nu$ would involve the treatment of Feynman diagrams with two loops. However, since KSVZ's $C_{\ax \tilde{f}f}$ is smaller than those in DFSZ models by two orders of magnitude, one can naively deduce that the decay amplitude of $\ax \to \gamma + \nu$ for KSVZ models is much smaller than those in DFSZ models. Henceforth, for simplicity, our focus is solely on axino DM in DFSZ models. 

We assume that the right-handed sfermion contributes dominantly to the decay amplitude of $\ax \to \gamma + \nu$. Summing Feynman diagrams (\ref{sf:a}) and (\ref{sf:b}) in Figure~\ref{fig:feyn}, the decay amplitude is found to be of the form $\sigma^{\mu\nu}k_{\mu}\epsilon^*_{\nu}$, where $k_{\mu}$ and $\epsilon_{\nu}$ are the momentum and the polarization vector of the photon respectively, as expected from gauge invariance. Considering only one trilinear RPV coupling at a time, and letting $\lambda$ denote $\lambda_{ijk}$ and $\lambda_{ijk} ^{\prime}$, the decay rate reads
\begin{eqnarray}
\Gamma(\ax \to \gamma + {\nu})=\frac{\lambda^2 e^2 C_{\ax \tilde{f}_Rf}^2 m_f^2 m_{\ax}^3}{4096\pi^5}\left|C_0(m_f,m_{\tilde{f}_R})\right|^2,
\label{eq:dy}
\end{eqnarray}
where $m_f$ and $m_{\tilde{f}_R}$ are the fermion and right-handed sfermion masses respectively.  $C_0(m_f,m_{\tilde{f_R}})$ is a loop function defined as follows:
\begin{eqnarray}
C_0(m_f,m_{\tilde{f}_R})=\frac{1}{i\pi^2}\int \frac{d^4q}{\left(q^2-m_f\right)^2\left((q-p)^2-m_{\tilde{f}_R}\right)^2\left((q-k)^2-m_f\right)^2},
\end{eqnarray}
where $p^{\mu}$ is the four-momentum of the axino, and $(p-k)^{\mu}$ is the four-momentum of the neutrino. $C_0(m_f,m_{\tilde{f}_R})$ scales roughly as $1/m^2_{\tilde{f}_R}$, and its mass dimension is inverse squared. 

The decay rate in Eq.~(\ref{eq:dy}) is proportional to $m_f^2$. This is due to a chirality flip at the fermionic internal line of the loop diagrams in Figure~\ref{fig:feyn}. Thus, loop contributions from (s)fermions of the third generation of (s)lepton or (s)quark have the most significant impact on the decay rate. The decay rate of the conjugate process $\ax \to \gamma + \bar{\nu}$ is the same, and gives a factor of two to the overall decay rate. Note that the overall decay rate must also be multiplied by the color factor for colored (s)fermions. 

So far, we have not discussed the axino-fermion-sfermion coupling from the Yukawa interaction, i.e.
\begin{eqnarray}
\Delta\mathcal{L}_{\ax \tilde{f}f}= C_{\ax \tilde{f}_Rf_L}\tilde{f}_R^*\bar{\ax}P_Lf +  C_{\ax \tilde{f}_Lf_R}\tilde{f}_L^*\bar{\ax}P_Rf^c + {\rm h.c.}\ ,
\label{eq:higgs}
 \end{eqnarray}
where $C_{\ax \tilde{f}_Rf_L}$ is the product of the axino-Higgsino mixing ($\sim v/f_a$) and the SM Yukawa coupling. The strength of $C_{\ax \tilde{f}_Rf_L}$'s for the third generation of fermions is comparable with those from the gauge interaction as in  Eq.~(\ref{eq:dfsz}). However, in contrast to the axino-fermion-sfermion coupling from the gauge interaction, i.e. Eq.~(\ref{eq:gauge}), where the decay amplitude picks up a chirality-flipping $m_f$ (see Eq.~(\ref{eq:dy})), the decay amplitude for contributions from the Lagrangian of Eq.~(\ref{eq:higgs}) picks up an $m_{\ax}$ instead~\cite{Mukhopadhyaya:1999gy}. Since $m_{\ax}\ll m_f$ for the third generation of fermions, we can neglect the contributions from the Lagrangian of Eq.~(\ref{eq:higgs}).

To be more concrete, let us consider $LLE$ RPV operators with (s)tau running in the loop ($\lambda_{i33}, i = 1, 2$). From Eqs.~(\ref{eq:dfsz}) and~(\ref{eq:dy}), the lifetime of the axino reads
\begin{eqnarray}
   \tau_{\ax}\simeq 8.7\times 10^{27}~{\rm sec} \lrfp{\lambda_{i33} \hat{g}}{10^{-4}}{-2} \lrfp{m_{\ax}}{7~\KEV}{-3} \lrfp{f_a/v}{10^{8}}{2}\lrfp{m_{\tilde{\tau}_R}}{100~\GEV}{4}\lrfp{m_{\tau}}{1.77~\GEV}{-2}.
   \label{eq:nm}
\end{eqnarray}
Here, we have factored out the dimensionful parameter $1/m^4_{\tilde{\tau_R}}$ from $\left|C_0(m_f,m_{\tilde{f}_R})\right|^2$, and used $\left|C_0(1.77~\GEV,100~\GEV)\right|\simeq7 \times 10^{-4}~\GEV^{-2}$. It can be seen from  Eq.~(\ref{eq:nm}) that $f_a \sim {\cal O}(10^{9}-10^{11})~\GEV$ for $\lambda_{i33}\sim {\cal O} (10^{-3}-10^{-1})$. This parameter region is consistent with the observed 3.5 keV X-ray line.

\subsection{Constraints}
Next, we study possible experimental constraints on $\lambda_{i33}$. Trilinear RPVs generate neutrino masses at the one-loop level. The neutrino mass matrix ($M^{\nu}_{ij}$) receives the following contributions from RPV couplings $\lambda_{ijk}$~\cite{Hall:1983id,Babu:1989px,Barbier:2004ez}:
\begin{eqnarray}
M^{\nu}_{ij}=\frac{1}{16\pi^2}\sum_{k,l,m} \lambda_{ikl}\lambda_{jmk}m_{e_k}\frac{({\tilde{m}_{LR}}^{e 2})_{ml}}{m^2_{\tilde{e}_{Rl}}-m^2_{\tilde{e}_{Lm}}}{\rm log}\left(\frac{m^2_{\tilde{e}_{Rl}}}{m^2_{\tilde{e}_{Lm}}}\right)+ (i \leftrightarrow j),
\end{eqnarray}
where the indices denote lepton's generation. $({\tilde{m}_{LR}}^{e 2})_{ml}$ is the left-right mixing matrix of slepton. We focus on the (s)tau loop decay scenario as discussed previously. Setting the related SUSY mass parameters to a common mass scale $m_{\rm SUSY}$, as done in~\cite{Barbier:2004ez}, the upper limit of neutrino mass, $m_{\nu}\lesssim 1~{\rm eV}$ implies that $\lambda_{i33} \lesssim 2 \times 10^{-3} (m_{\rm SUSY}/100 \GEV)^{1/2}$. This bound on the neutrino masses can be alleviated by raising the left-handed stau mass. For instance, when $m_{\tilde{\tau}_{L}}=5m_{\tilde{\tau}_{R}}=500~\GEV$, the bound is relaxed to $\lambda_{i33} \lesssim 0.02$.

We now switch our attention to other experimental constraints on $\lambda_{ijk}$. $\lambda_{133}, \lambda_{233}$ are constrained to be $\lambda_{133},\lambda_{233}\lesssim 0.07 \times (m_{\tau_R}/100~\GEV)$, which is derived from the measurements of $\Gamma(\tau\to\mu\nu\bar{\nu})/\Gamma(\mu\to e\nu\bar{\nu})$~\cite{Allanach:1999ic,Barbier:2004ez}.         

In addition to $\lambda_{i33}$, $\lambda_{i33}^\prime$ could reproduce the observed X-ray line signal as well. However, this means that sbottom has to be rather light, i.e. $\sim {\cal O}(100)~\GEV$ as in Eq.~(\ref{eq:nm}). Since the LHC is capable of producing colored particles copiously and imposing stringent mass bounds on colored particles ($\gtrsim 100~\GEV$ in general), (s)bottom-loop decay scenario is not favored.~\footnote{Sbottom decays into a quark and a neutrino via the trilinear RPV couplings. This mimics the $\tilde{q}\to q \tilde{\chi}^0$ and $\tilde{b}\to b \tilde{\chi}^0$ channels studied at the LHC, where $\tilde{\chi}^0$ is the lightest neutralino~\cite{atlas-sq,Aad:2013ija}. The bound on the squark mass is 780 GeV at 95\% CL when $m_{\tilde{\chi}^0} \to 0~\GEV$ and gluino is decoupled~\cite{atlas-sq}. For sbottom, the bound is 620 GeV at 95\% CL for $m_{\tilde{\chi}^0} < 120~\GEV$~\cite{Aad:2013ija}.}

Before closing this Section, let us comment on the possibility of explaining the 3.5 keV X-ray line with gravitino DM, which is also a well-motivated decaying dark matter candidate in SUSY~\cite{Takayama:2000uz,Buchmuller:2007ui}. Within the bilinear RPV scenario, the lifetime of gravitino is
\begin{eqnarray}
\Gamma (\psi_{3/2} \to \gamma + \nu) \sim 2 \times 10^{46}~{\rm sec} \lrfp{m_{3/2}}{7~\KEV}{-3}\lrfp{\kappa_i}{10^{-7}}{-2},
\end{eqnarray}
where $m_{3/2}$ is the gravitino mass, and we have assumed that the photino-zino mixing parameter takes the value $|U_{\gamma Z}|\sim 0.1$. It is clear that the lifetime is far too long to be recognized as the source of the 3.5 keV X-ray line. This is due to the suppression from the enormous Planck scale, $M_{\rm Pl} = 2.4 \times 10^{18} \GEV$, as well as the tiny bilinear RPV couplings $\kappa_i$'s, which have tree-level contributions to neutrino masses; their values are stringently constrained~\cite{Romao:1999up,Takayama:1999pc,Hirsch:2004he}. In fact, the smallness of $\kappa_i$'s is also the reason why it is difficult to explain the X-ray line using axino DM with bilinear RPV. Returning to the discussion of gravitino DM, let us note that in certain situation, gravitino DM with trilinear RPV can decay dominantly into photon and neutrino~\cite{Lola:2007rw,Liew:2013mta}. The details of this possibility are out of the reach of this work and are left for future studies.~\footnote{Upon completing this work, we found that 7 keV gravitino DM is also discussed in~\cite{Kolda:2014ppa}, overlapping with arguments given here.}

\section{Cosmological and phenomenological implications}
\subsection{Cosmology}
Late decaying next-to-the-lightest MSSM SUSY particle (such as neutralino, stau or sneutrino) could spoil the success of big-bang nucleosynthesis (BBN), which is well determined within the SM. However, for the parameter region of $\lambda$ in consideration, such particle decays well before the BBN begins, keeping standard BBN processes intact~\cite{Gherghetta:1998tq,Takayama:2000uz,Buchmuller:2007ui}. 

In the early universe after reheating, axino is produced via thermal interactions of SM and SUSY particles. Two classes of interactions are particularly important: anomaly-induced terms as in Eq.~(\ref{eq:anom}), and terms from the the superpotential $W_{\rm PQ}= k \Phi\Psi\bar{\Psi}$, where $\Psi$'s are replaced by the heavy quark (Higgs doublet) superfields in KSVZ (DFSZ) models. $\Phi$ is the PQ superfield. In DFSZ models, the mass of $\Psi$, $M_{\Psi}$ is around the weak scale, and Yukawa interactions from $W_{\rm PQ}= k \Phi\Psi\bar{\Psi}$ are important to the thermal production of axino as long as the reheating temperature $T_{\rm R}$ is larger than the weak scale. Assuming that this contribution is dominant, the relic abundance of axino is found to be fairly independent of $T_{\rm R}$~\cite{Chun:2011zd,Bae:2011jb,Choi:2011yf,Bae:2011iw}. For $M_{\Psi}= 1$ TeV and $T_R \gtrsim 10^4~\GEV$, the abundance reads~\cite{Bae:2011jb}
\begin{equation}
\Omega_{\ax}^{\Phi\Psi\bar{\Psi}} h^2 \sim {\cal O}(1)\times \left( \frac{m_{\ax}}{10 \,{\rm keV}} \right)\left( \frac{f_a}{10^{10}\,{\rm GeV}} \right)^{-2},
\end{equation}
indicating that within the parameter region of interest, thermal abundance of axino can account for the main component of DM. Relic abundance coming from the anomaly-induced terms is proportional to $T_R$ (see \cite{Brandenburg:2004du,Strumia:2010aa} for a proper treatment), but it is relatively small up to $T_R\simeq 10^6-10^7~{\rm GeV}$~\cite{Choi:2011yf}.~\footnote{It is however argued in~\cite{Bae:2011jb} that this contribution is suppressed even at higher $T_R$.} Therefore, $T_R\sim 10^4~{\rm GeV}$ is the suitable value of reheating temperature in our scenario. Note that there could be other factors affecting the axino abundance, such as the decay of moduli or massive particles, either increasing the number of axino by direct decay, or reducing its number density via dilution. These situations are possible but highly model-dependent. 
 
The abundance of axion from coherent oscillation with initial misalignment angle $\theta_a$ is determined as~\cite{Turner:1985si}
\begin{equation}
	\Omega_a h^2 \simeq 8\times 10^{-4} \theta_a^2 \left( \frac{f_a}{10^{10}\,{\rm GeV}} \right)^{1.19}.
\end{equation}
For $f_a \sim {\cal O}(10^{9}-10^{11})$ GeV, the axion abundance in the present universe is negligible as far as $\theta_a \lesssim {\cal O}(1)$.

The mass of gravitino can be much heavier than axino, depending on the underlying SUSY model~\cite{Kim:1983ia,Moxhay:1984am,Goto:1991gq,Chun:1992zk,Chun:1995hc}. If gravitino is the next-to-the-lightest SUSY particle, it will mainly decay into an axion and an axino: $\psi_{3/2}\to \ax + a$ without causing cosmological gravitino problem since the decay products interact weakly with other particles~\cite{Asaka:2000ew}. Gravitino is produced thermally in the early universe mainly via scatterings with gluon and gluino, and its relic abundance reads~\cite{Moroi:1993mb,Bolz:2000fu,Pradler:2006qh,Rychkov:2007uq}
 \begin{equation}
\Omega^{(\rm th)}_{3/2}h^2\simeq 0.03\left( \frac{T_R}{10^{4}\,\rm{GeV}}\right) \left( \frac{m_{3/2}}{1\,\rm{MeV}}\right)^{-1} \left( \frac{m_{\tilde{g}}}{1\,\rm{TeV}}\right) ^2, 
\label{eq:gr}
\end{equation}
where $m_{\tilde{g}}$ is the gluino mass. By requiring $\Omega^{(\rm th)}_{3/2}h^2\lesssim 0.1$ and $T_{\rm R} \gtrsim 10^4~\GEV$, we can derive the lower bound on the gravitino mass: 
$m_{3/2}\gtrsim 300$ keV for $m_{\tilde{g}} \simeq 1$ TeV. Let us briefly comment on the scalar partner of axion, namely saxion. The saxion mass is expected to be $m_{\sigma}\sim m_{3/2}$, and the detailed physics of saxion is quite model-dependent. For $m_{\sigma}\gtrsim 100$ keV, cosmological constraints can be satisfied in general in the region of reheating temperature of concern~\cite{Kawasaki:2007mk}.  

Baryon asymmetry generated in the early universe could be washed out by $B-L$-violating RPV interactions. However, this problem is evaded as long as one of the lepton flavors of the RPV interactions is small enough such that it decouples from the thermal bath sufficiently early~\cite{Campbell:1990fa,Fischler:1990gn,Dreiner:1992vm,Endo:2009cv}. Details of this effect are dependent on the flavor structure of the theory.

Finally, let us mention the impact of the recent BICEP2 result on our model~\cite{Ade:2014xna}. If the reported detection of primordial B-mode polarization is true, the case where the PQ symmetry breaking occurs before or during inflation is stringently constrained by the isocurvature perturbations. One way to avoid this is by considering post-inflationary PQ symmetry breaking. However, in this scenario, the DFSZ axion model discussed here would lead to the cosmological domain wall problem. This problem is resolved by the combined KSVZ-DFSZ models where the domain wall number is restricted to unity.        
  
\subsection{Particle phenomenology}
We emphasize that due to the fermion mass suppression on the axino decay rate (Eq.~(\ref{eq:dy})), The RPV couplings relevant to the first and second generations of leptons have negligible effects on the axino decay rate. Moreover, within the parameter region of interest, i.e. $f_a \sim {\cal O}(10^{9}-10^{11})~\GEV$, bilinear RPV interactions do not have significant contributions to the decay rate of axino~\cite{Kong:2014gea}. This means that one can reconstruct the neutrino mass matrix using these RPV parameters without affecting the decay rate of axino. For example, assuming that the bilinear RPV effects on the neutrino mass matrix are dominant, it is possible to generate the atmospheric neutrino mass scale at the tree level from bilinear RPV couplings. Other entries of the neutrino mass matrix are generated via radiative corrections~\cite{Hirsch:2004he}.

 We now briefly discuss the collider phenomenology of this model, focusing on the consequences of RPV interactions. Collider physics of SUSY with RPV is an important topic, and there are numerous references on it in the literature. Here, we are in particular interested in the prospects of studying light (right-handed) stau, as required by our axino DM scenario. The dominant decay mode of the stau is: $\tilde{\tau} \to l + \nu$. This decay channel resembles the slepton-decays-into-neutralino channel: $\tilde{l} \to l + \tilde{\chi}^0$ with $m_{\tilde{\chi}^0} \to 0$.
 
 Let us consider the production of stau at colliders. Lepton colliders such as the ILC have especially bright prospects of looking for the direct production of staus. At hadron colliders, stau signals can also be looked for via cascade decays of SUSY particles. For example, in~\cite{Desch:2010gi}, the following cascade process at the LHC has been considered: $pp \to \tilde{q}\tilde{q}\to jj\tilde{\chi}^0\tilde{\chi}^0\to jj\tau\tau\tilde{\tau}\tilde{\tau}$, where $j$ refers to jet. In this situation, $\tilde{\tau}$ is assumed to be the lightest MSSM SUSY particle. It is seen that the decay signals  of stau are accompanied by hadronic jets.

\section{Conclusions}
In the present work, we have considered axino DM with trilinear RPV in the wake of an anomalous X-ray line found to be originated from galaxy clusters and Andromeda galaxy. We have found several interesting features within this framework, including a consistent interpretation of the line signal with the phenomenologically viable ``axion window" ($f_a \sim{\cal O}(10^{9}-10^{11})~\GEV$), as well as a light stau ($\sim{\cal O}(100)$ GeV). Cosmological constraints can also be satisfied in general. The next run of the LHC will be crucial at identifying or disfavoring the model by searching for the characteristic R-parity violating decay of stau. Finally, let us remark that it is also vital to obtain experimental confirmation of the tentative line signal from other X-ray telescopes, such as the forthcoming ASTRO-H. 

\section*{Acknowledgment}
The author is grateful to K.~Hamaguchi and K.~Nakayama for fruitful discussions and useful suggestions. This work was supported by the Program for Leading Graduate Schools, MEXT, Japan.

\end{document}